\documentclass[12pt]{article}

\usepackage{natbib}
\setcitestyle{authoryear,round,aysep={}}

\usepackage[utf8]{inputenc}

\usepackage{color}
\usepackage{placeins}

\usepackage[dvips]{graphicx}

\usepackage[dvips]{graphics}

\usepackage{amssymb}

\usepackage{amsmath}
\usepackage{epsfig}
\usepackage{epsf}

\begin{document}

\title
{\large {From anti-conformism to extremism}}

\author{
 G{\'e}rard~Weisbuch\\
Laboratoire de Physique Statistique\\
et Centre de Recherche sur l'Environnement et la Soci{\'e}t{\'e}\\
     de l'Ecole Normale Sup{\'e}rieure, \\      
    24 rue Lhomond, F 75231 Paris Cedex 5, France. \\
  {\small     {\em email}:weisbuch@lps.ens.fr}\\ 
 }
\maketitle

\begin{abstract}

  We here present a model of the dynamics of extremism based on opinion
dynamics in order to understand the circumstances which favour its
emergence and development in large fractions of the general public.
Our model is based on the bounded confidence hypothesis and on the 
evolution of initially anti-conformist agents to extreme positions. 
  Numerical analyses demonstrate that a few anti-conformist are able to drag
a large fraction of conformists agents to their position provided
that they express their views more often than the conformists.
The most influential parameter controlling the outcome of the
dynamics is the uncertainty of the conformist agents; 
the higher their uncertainty, the higher is the influence of anti-conformists.
Systematic scans of the parameter space show the existence of two
regime transitions, one following the conformists uncertainty parameter
 and the other one following the anti-conformism strength.

\noindent


\end{abstract}

\section{Introduction}

  The present paper discusses the dynamics of extremism
in a democratic setting. A probably over-optimistic view
of democracy is that when opinions are openly expressed,
some consensus opinion would emerge and citizens would vote 
in favour of a government whose actions would be in accordance 
with the views of a large majority of citizens. This utopia
is shared by many writers, but 
History has consistently shown us that National Consensus was
a dream, that could eventually occur at war time, not such a
wishful situation.

  At least one would expect that the elected government would
be close enough to a centrist position satisfying the largest
proportion of citizens, as the ice cream seller choosing
to put his stand near the middle of a linear beach (\cite{hotel}).

 Once again, History since the Eighteenth century Enlightenment period
in Western Europe contradicts these simple views and we are not observing 
a smooth evolution towards more consensus nor towards the success
of centrists parties. We rather observed alternation
between regimes of dominance of centrists political parties and regimes of strong
ideological fights between more extremist parties, eventually leading to
de facto dictatorship, according to time periods and world regions. 
  The present paper is an essay to model possible evolutions
of public opinions leading to different opinion aggregation landscape 
forming the basis of political entities corresponding to parties.
We here develop a model of opinion dynamics in order
to answer such questions as:
\begin{itemize}
\item How come rational\footnote{rational here does not refer 
to economists' full rationality but rather to its common sense, people able 
to practice some form of reasoning} 
people choose extremism?
\item How does an initial low proportion of anti-conformist 
influences/(or does not), a large fraction of the general population to
aggregate in powerful extremist clusters?
\item What characterises political clusters in terms of 
the number of agents in the cluster and their distance to
a middle opinion? More precisely
what are the regions in the parameter space of the model which would
lead the different outcome of the dynamics? 
\end{itemize}

  The simulations presented in this paper are based on opinion dynamics: 
agents exchange their views on the occasion of encounters,
and they might update their opinion as a result of these exchanges.  
 We are well aware that opinion formation in politics involves 
many other processes than encounters and discussions among individuals: 
media, political parties, the government and other political
institutions are involved as well. For the sake of clarity,
we postpone the discussion of the robustness
of our results with respect to these other factors to the 
last section of the paper.

The earliest models of opinion dynamics 
were binary opinions models, where opinions could take only
two discrete values, e.g. -1 and +1 in the so-called voters models as described 
in \cite{holley75},\cite{galam82} and summarised in \cite{rmp}.

We here present a model based on continuous opinions,
more adapted to the discussion of the assets and liabilities
of political choices among agents, and to the traditional right/left
axis of political analysts, than binary opinions.  
It is inspired from two approaches,
the bounded confidence model of \cite{dna}
 and the anti-conformism model of
 \cite{smalep}. Since these two models are used as building blocks
of our model, we will first summarise their main aspects.
 
  The rest of the paper is then divided into 3 sections

\begin{itemize}
\item Short reminders of the previous models.
\begin{itemize}
\item \cite{dna} bounded confidence model including its
application to extremism.
\item Smaldino and Epstein model of anti-conformism.
\end{itemize}
\item  Our synthetic model is developed and its results presented.
\item  Conclusions and discussion.
\end{itemize}
 
Disclaimer

  The present paper should not be interpreted as normative: 
we rather try to describe the evolution of opinions and political choices.
One can certainly give examples such as Civil Rights in general, when initially
considered extremist opinions were later largely accepted by the public.
And other cases in which the consequences of extremism turned out
to be dramatic.    

\section{Essentials of former models}

  In order to achieve consistency in notations and hypotheses,
we use our own notation throughout the paper, which sometimes differ from
those of \cite{dna} and \cite{smalep} and make appropriate scale changes.

\subsection{Bounded confidence}

  The bounded confidence model is based on a major
cognitive bias, the confirmation bias (\cite{plous}): we are mostly influenced
by opinions close to ours and tend to reject opinions too
far away. The mathematical model was independently introduced by \cite{dna} and  by
\cite{HK}. It follows the spirit of Axelrod's earlier model
 of dissemination of cultures.
In \cite{axel} model, cultures are described by strings of integers.
Pairs of agents interact if their cultures are already close enough,
in which case one of them adjusts 
one feature of its culture string to match that of the other agent's culture.

In bounded confidence models, opinions are represented by real numbers. When opinion
differences are lower than a confidence threshold, agents adjust their opinion 
by decreasing such difference. In Deffuant et al. model, pairs of agents are randomly
chosen in the population of agents and they eventually adjust their opinion if the 
confidence condition is met. Another pair is randomly chosen, and so on.
 Such an iteration mode is called random sequential.

 By contrast \cite{HK} apply the same
opinion updating equation but they use parallel iteration:
 all opinions are updated simultaneously. Their choice is well
adapted to discussions in committees for instance.

We will consistently use random sequential iteration in this paper.     

\subsection{Deffuant et al. bounded confidence model}
 
  \cite{dna} bounded confidence model was introduced to model
situations in which actors have to take decisions involving cost/benefit
analysis in terms of money. Such was the case when the Common Agricultural Policy was
modified in 1992: farmers were proposed to change their former practices in favour
of more environment friendly practices, e.g. by reducing 
fertilisers and pesticides use, in exchange for financial aid.
 But optimising new practices involved a lot of 
financial uncertainties and surveys demonstrated that farmers would have many
social interactions discussing the pros and the cons of the environmental contracts
before taking any decision.

\cite{dna} model can be simply described:

Opinions are
represented by a {\bf continuous variable} $x$.

Two randomly chosen agents with opinions $x$ and $x'$
interact if, and only if, $|x-x'|<u$. Opinions are updated according to: 
\begin{equation}
\begin{array}{c}
  x = x + \mu \cdot (x'-x) \\
  x' = x' + \mu \cdot (x-x') 
\end{array}
\end{equation}

   $u$ represents the uncertainty of the agents,
 and $\mu$, taken between 0 and $0.5$,
is a kinetic parameter. If the two initial opinions are close enough, the two
agents interact and their opinions move closer. Otherwise, no opinion change occurs\footnote{
Many extensions of the bounded confidence model were proposed 
as described in the review of \cite{rmp}.
Some take into account the possiblity of repulsion among agents such \cite{huet,aliza}:
agents can be either attracted for small differences in opinions,
 but can also have repulsive interaction
when their difference is larger than another upper thresold. 
Other models, \cite{kurmy}, consider two populations of interacting agents, some
having only attractive interaction, others also having repulsive interactions.}.

\begin{figure}[!h]
\centerline{\epsfxsize=120mm\epsfbox{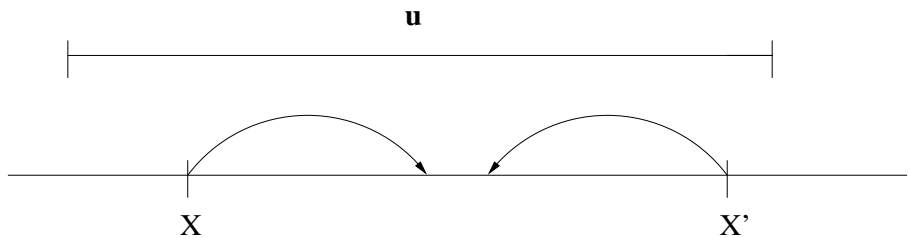}} 
\caption {Agents with initial positions $x$ and $x'$ move there opinion closer
to each other opinion.
 The threshold for actual interaction $u$ is interpreted as a confidence
  or uncertainty parameter.}
\end{figure}

\paragraph{Simulations}  

  An initial distribution of agents opinions is first randomly established.
To achieve maximum randomness most authors choose a uniform distribution on
a segment. The first model of Deffuant used [0,1]
as initial segment. Later most authors used  [-1,1] as initial segment
which we do in this paper for the sake of comparison.

At each time step, a random pair of agents is chosen, to which the above described 
updating algorithm is applied.

Simulations are stopped after convergence of opinions into one or several clusters.

\paragraph{Results}

  Opinions vs time plots, figure 2, represent opinion dynamics.
Each point on the graph represents the opinion of an individual agent along the y axis
 at time t on the horizontal axis when the agent is tested for opinion change.
 The time unit for all plots corresponds to 1000 pair updatings
 (on average each agent is tested twice per time unit).
 Individual dots might hardly be distinguished 
on these plots, but the envelope of the clouds gives an indication of the gradual convergence
of opinions in the course of time. 

\begin{figure}[!h]
\centerline{\epsfxsize=80mm \epsfbox{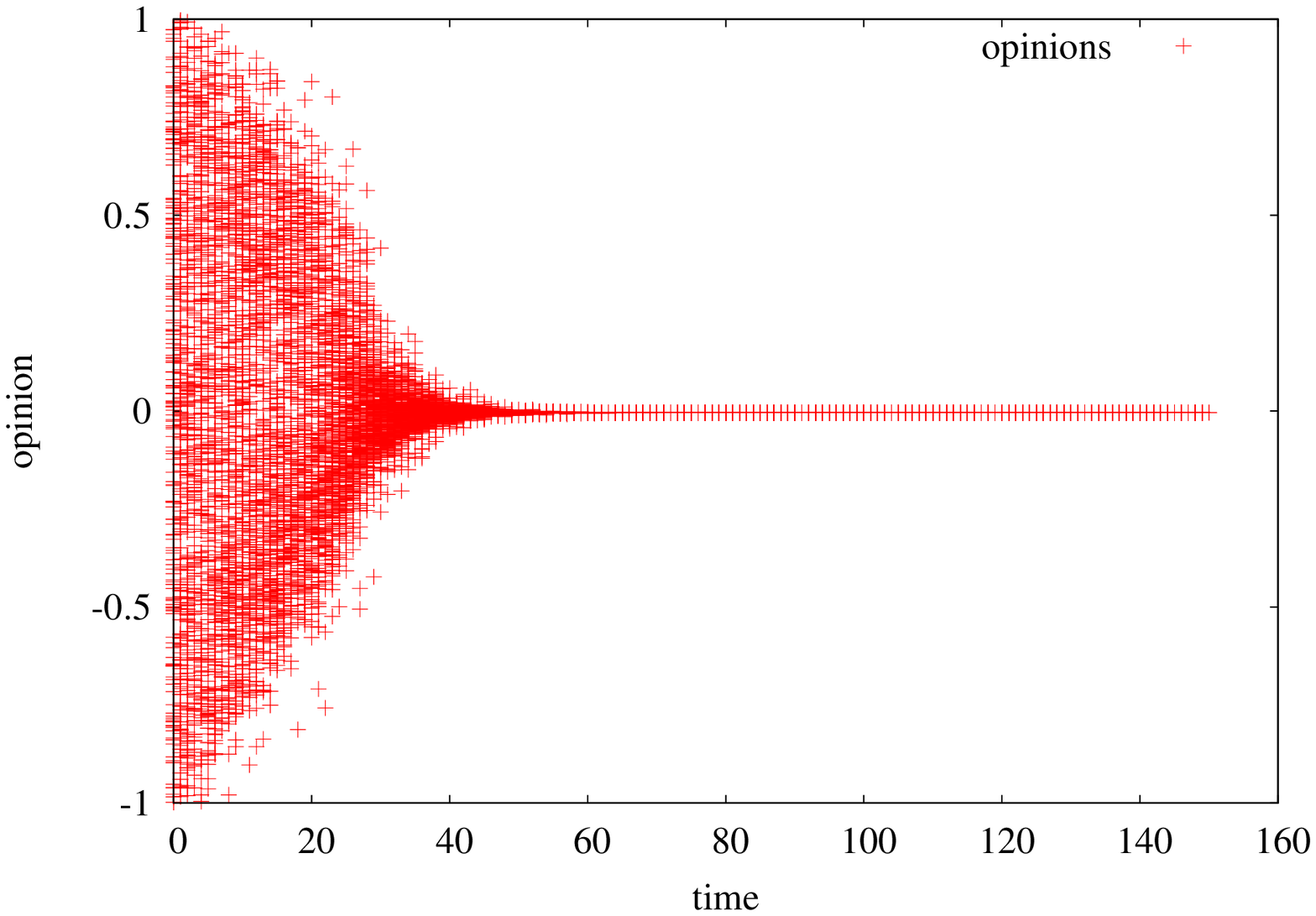} \epsfxsize=80mm \epsfbox{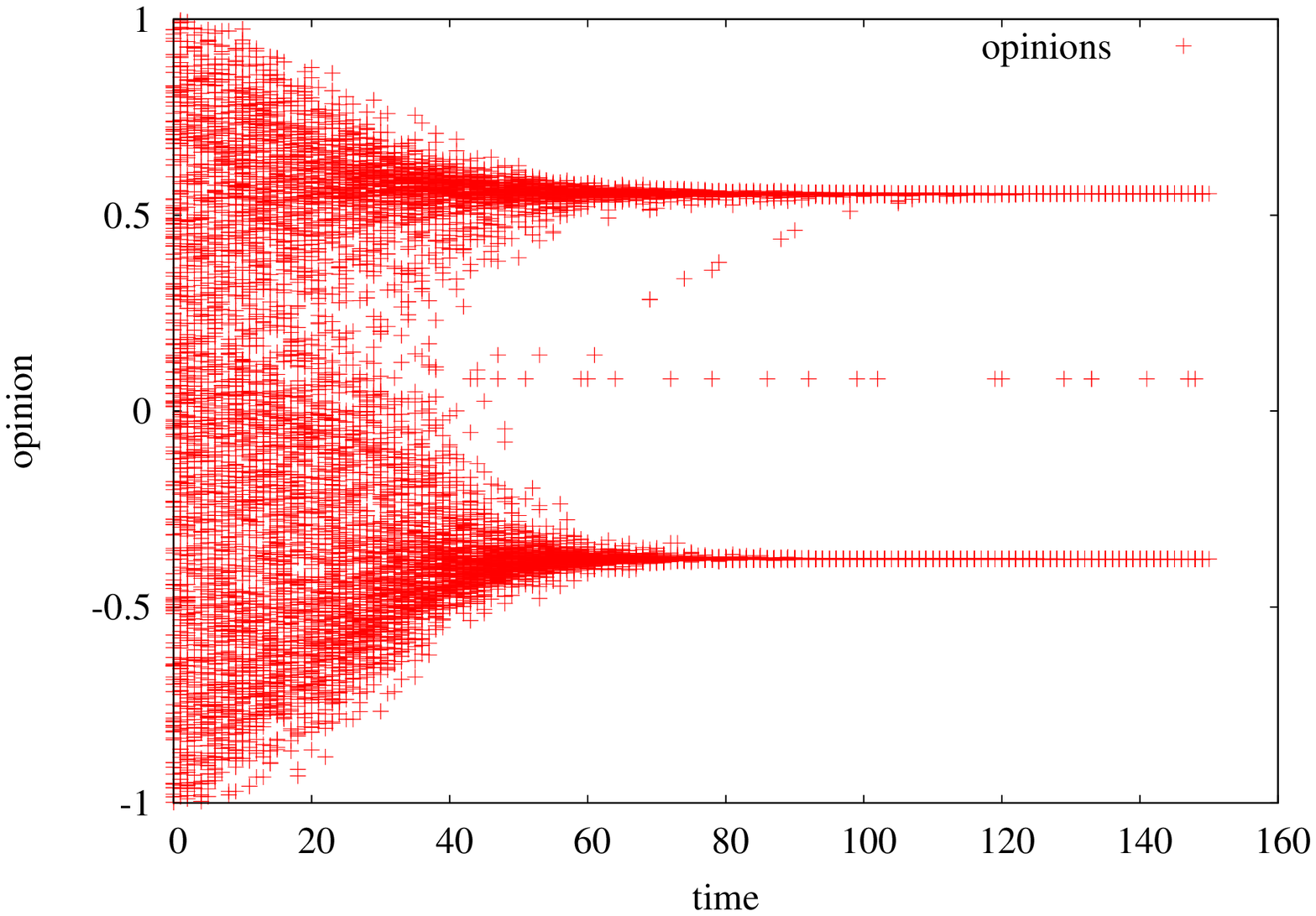}}
 \caption{Comparison of opinion dynamics for different uncertainties $u$.
 ($u=0.6$ for the left plot,
 $u=0.4$ for the right plot).}
\end{figure}

These two plots compare the evolution of opinions until convergence.
The number of agents is 1000, $\mu=0.1$, initial opinion uniformly and randomly
chosen on segment [-1,1]. Uncertainty $u$ is 0.6 for the left plot 
and 0.4 for the right plot.
Time 150 correspond to sampling 150 pairs, so each agent has been sampled for updating
300 times on average.   

\cite{dna} have shown that agents uncertainty $u$
 is the main determinant of the outcome of the dynamics. 

 To summarise the results relevant to our analysis:
\begin{itemize}
\item Opinions are clustered in $n=int(\frac{1}{u})$ clusters which do not interact
any more after a long enough time ($int$ stands for integer part).
 \item Clusters are of equal size and are at least $2u$ apart.
\end{itemize}
  We will further refer to the above statements as the $n=int(\frac{1}{u})$ rule.
The general expression for an initial distribution of opinions on
a segment of width $w$ is $n=int(\frac{w}{2u})$.

\FloatBarrier

\subsection{Deffuant et al. model of extremism}

\cite{daw} later proposed of model of extremism prevalence
inspired from the bounded confidence model. Their extremism model bears
certain differences with the bounded confidence model, the most relevant
for us is the introduction of a small fraction of extremists in the agent population.
Extremists are different from the other agents by having a very low uncertainty 
and by having extreme opinions at the end of the spectrum $[-1,1]$.
\cite{daw} also added a dynamics on uncertainty: agents not only exchange
 opinions but also uncertainty $u$. 

The main issue in \cite{daw} is whether the small fraction of extremists
is able to drag towards extremism the normal (centrist) agents
 with initially larger uncertainty and extended
 distribution of opinions. 

The initial fraction of extremists plays some role in the outcome of the dynamics, 
but the most significant result is that the centrists uncertainty $u$
determines the outcome of the dynamics in large regions of parameters.
For narrow uncertainties, say $u \leq 0.4$, a few perc. of centrists are dragged
toward extremism. For $u \simeq 1$ initially moderate agents
 are split in two opposed extremist fractions,
and for $u \geq 1.4$ most moderate agents are dragged towards
an asymmetric single extremist attractor, close to either -1 or +1.      

   This extremism model and the full set of results described in \cite{daw},
and more recently in \cite{def2006},
show that extremists can convert a large fraction of the population
for the largest values of the centrist uncertainty, even when their number
is relatively small with respect to the total population. What is missing is
the reason why extremism first arises. \cite{daw} postulate the initial existence
of some extremists in the population from empirical observations of the political
scene of most countries, including democracies.
The model by  \cite{smalep} provides a clue to the origin
of extremism.

\subsection{Anti-conformism generates extremism}

  \cite{smalep} recently introduced a model of 
anti-conformism and its consequences on individual preferences:
``Social Conformity Despite Individual Preferences for Distinctiveness''.
\footnote{ Their paper does not explicitely refer to politics but they quote several references
about politics.} 
   The concept of distinctiveness plays a central
role in several social psychological theories of self and identity processes.
In such framework, positions are most often taken in multi-dimensional spaces.
 Since we here represent political opinions as continuous variables
on a bounded support, the most distinctive opinions should be those at the boundaries. 
The idea is that some political agents choose anti-conformist attitudes
 to attract attention and get some prestige. The process can be observed
 among political activists and we will further discuss in the conclusion
 section why some professional political agents choose non-conformism or extreme positions.

 In the Smaldino-Epstein model, instead of exchanging with other agents
to share common views, anti-conformist agents react to the distribution of opinions
 (which they are supposed to be aware of) and they choose opinions
away from the average opinion of the other agents.
 Anti-conformists view as ideal
a position $x^*$ such that:

\begin{equation}
   x^* = x_{aver} + \delta \cdot \sigma
\end{equation}
 where $x_{aver}$ is the average opinion of the distribution, $\sigma$
is the standard deviation of the distribution and $\delta$ a kind
 of anti-conformist strength.
Agents then gradually update their opinion in the direction of $x^*$ according to:

\begin{equation}
   x = x + \mu \cdot (x^*-x)
\end{equation}

  Although anti-conformists wish to remain distinct from the crowd,
since they all share the same goal $x^*$ the variance of the distribution 
actually decreases during each iteration by a factor $1-\mu$
and they converge towards a single
attractor. Starting from an initial uniform distribution on
segment [-1,+1],
for positive values of $\delta$, the final opinion cluster 
is well above the initial opinion average at
  $\sigma_0\delta$, where $\sigma_0$ is the initial standard deviation.
 The asymptotic opinion
 can then stand outside the initial opinion range for
large values of $\sigma_0\delta$.

 In other words, anti-conformism\footnote{Other authors have introduced
anti-conformist agents in the simulation of binary opinion dynamics
\cite{serge,andre}. In the context of binary opinions, say 0 or 1,
anti-conformists have opinions {\bf opposed} to the opinion of their neighbours.
In the \cite{smalep} model as in the present paper, the anti-conformists
choose opinions {\bf further} than those of other agents.
 Their position can be described
as 'plus royaliste que le Roi' or in English 'more catholic than the Pope'. 
 Hence, the dynamics of 'our' mixed population 
is quite different from those described in \cite{serge,andre}} results
in convergence to more extreme opinions than the initial average opinion, 
which makes the process a valuable hypothesis on the origin of extremism.
Of course many other factors can play a role, but we will here only investigate
the importance of the anti-conformism factor\footnote{The \cite{smalep}
paper covers more situations than reported here, including heterogeneity of  
$\delta$, the anti-conformist strength. It e.g. shows that 
the above conclusions on convergence of the dynamics
remain true provided that such heterogeneity is limited:
the standard deviation of the $\delta$ distribution should be less than 1.}.
\bigskip
\begin{figure}[h!]
\centerline{\epsfxsize=110mm\epsfbox{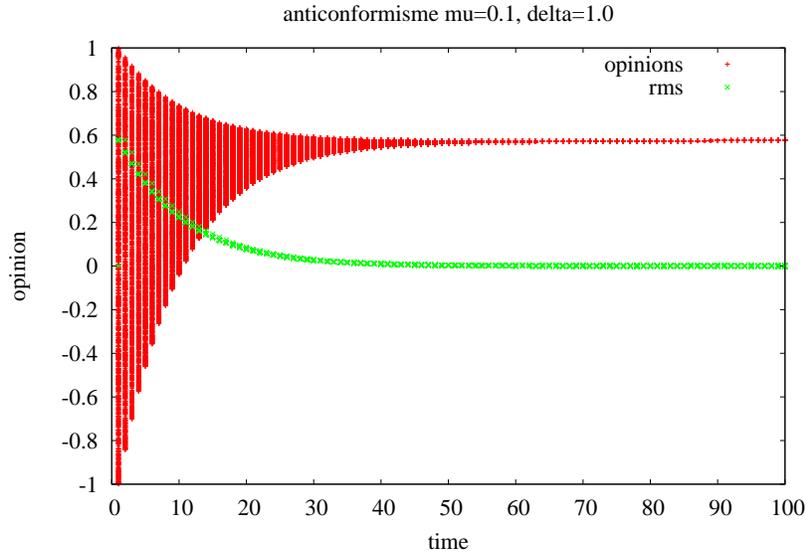}}
\caption{Anti-conformism generates extremism: 
evolution of opinions (red dots) and of the standard deviation (green dots)
for 1000 anti-conformist agents with kinetic parameter $\mu=0.1$
and anti-conformist strength $\delta=1$.}
\end{figure}

\FloatBarrier

\section{The synthetic model}

In the present synthetic model, two different populations of agents 
are introduced:
\begin{itemize}
\item Conformists obeying a bounded confidence model as in \cite{dna}. 
The conformists can be directly influenced by all individual 
agents within their uncertainty range. Their opinions aggregate into clusters.
\item Anti-conformists obeying the anti-conformist model of \cite{smalep}.
The anti-conformists are only influenced by the local properties, average
opinion and standard deviation of the agents' opinion, within their uncertainty range.
As a result, they soon move towards the exterior of the distribution of conformists.
\end{itemize}
  Numerical simulations show that a large fraction of the conformists agents 
also evolve towards extreme positions where they form extra clusters with anti-conformists.

  Let us stress the difference between the anti-conformists of the present model
and the extremists of \cite{daw}. In \cite{daw}, the extremists have been given
{\bf ab initio} extreme positions and they are far less susceptible to centrists 
attraction than the other agents. In the present model, anti-conformists
have the same initial opinion distribution than the other agents.
 They evolve towards extreme positions because of their specific dynamics. 
They become little influenced by centrists when they get close to their equilibrium
position because meanwhile centrists opinions have coalesced into clusters,
thus reducing the standard deviation of their distribution.
 In other words, anti-conformists {\bf acquired} a role similar to extremists.
By contrast, the attraction of conformists towards anti-conformist has been identically maintained.
   
\subsection{Model description}

Two populations of conformists and anti-conformists co-exist and interact 
according to the threshold condition.
For any agent, anti-conformist or conformist, interactions only involve their neighbourhood,
$[x-u,x+u]$. $u$ is the same for conformists and anti-conformists and
 specifies the interaction range for both types of agents.

We start from a uniform initial distribution of agents opinions
on the [-1,+1] segment.

At each time step, one agent is first randomly selected.

 If an anti-conformist were selected during the first draw,
it interacts with all the agents within the reach of its opinion
$[x_e-u,x_e+u]$ using the Smaldino-Epstein algorithm,
moving towards the local $x^*$. This happens with probability $r$
 where $r$ is the fraction of anti-conformist agents. 
 And that's it for the round.

If the first agent were a conformist, which happens with probability $1-r$,
 a second agent is randomly selected
and we apply the Deffuant algorithm. In fact, if the second agent
were an anti-conformist, only the conformist moves towards the anti-conformist
since anti-conformists are not involved in binary interaction\footnote{We
 have chosen not to move the anti-conformist,
 although the alternative choice, move according to SE rule
 could have been made.
 Anyway, differences in behaviour between
 the two choices would not have changed qualitatively
 the dynamics for our choice of parameters.}. In other word,
 their only moves are those dictated by the Smaldino-Epstein algorithm.

Figure 4 describes the tree of probabilities for one sampling round.

\bigskip
\begin{figure}[h!]
\centerline{\epsfxsize=80mm\epsfbox{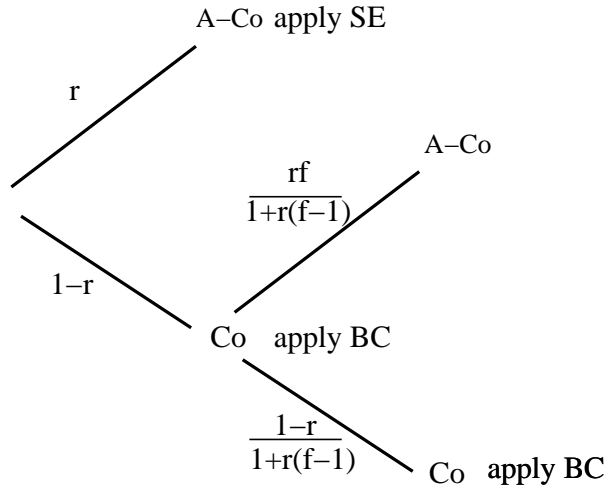}}
\caption{ The tree of probabilities for one iteration step.
  A-Co stands for anti-conformist agents and Co for conformists. 
The applied algorithms are noted SE (resp. BC) for Smaldino Epstein
 (resp. for Bounded confidence). Refer to the text for the
computation of probabilities of each branch.}
\end{figure}

\FloatBarrier

  Since we here describe the political arena, we further add two more 
rules for anti-conformists.

\begin{itemize}
\item 
 Anti-conformists choose positions outside the crowd
and as far as possible from the other party(ies); a simple implementation
of this inclination is to choose positive (resp. negative) 
values of $\delta$ for anti-conformists
with positive opinion (resp. negative). This simple rejection rule
is only applied to anti-conformists\footnote{This simple implementation
 does not cause
any practical problem for opinion values close to zero 
since anti-conformists move away very soon from
 the average as observed in figures 5 and 7.}.
 Anti-conformist opinions
are only updated when the first draw was an anti-conformist.
\item 
 We take into account the fact that anti-conformists are more active in propaganda
than conformists. They express their opinion more often than conformists, whether
in the streets or in the media (\cite{bron}). One extra parameter of the model is $f$
the relative frequency 
of opinion expression of anti-conformists with respect to the frequency of
expression of conformists.
This is implemented in the model by the fact that any anti-conformist is randomly selected
 for interaction $f$ times more often than any conformist.
\end{itemize}

Let us be more specific about the implementation of the second rule
concerning the second draw.
Let $r$ be the fraction of anti-conformist in the population.
Since any anti-conformist is selected $f$ more often than any conformist,
anti-conformists are selected for interaction with a probability proportional to  $rf$.
Conformists are selected with a probability proportional to $1-r$.
Normalising probabilities to sum 1 implies to draw anti-conformists with probability
$\frac{rf}{1+r(f-1)}$ and conformists with probability $\frac{1-r}{1+r(f-1)}$.
   
\subsection{Simulation results}

\paragraph{Time plots and histograms}

  Let us first compare time plots (fig. 4 and 6) and asymptotic histograms (fig. 5 and 7)
of opinion dynamics,
  when one changes the relative expression frequency parameter $f$ from 1 to 20.
The following plots were drawn for an uncertainty level $u=0.3$, 
anti-conformism strength $\delta=2$,
number of agents 1000, number of iterations per agent 300,
fraction of anti-conformists 0.05, kinetic factor $\mu=0.1$. 
Let us remind here for the sake of comparison that the 
 $int(\frac{1}{u})$ rule predicts 3 clusters at
opinions -0.66, 0 and 0.66 in the absence of anti-conformists.

\begin{figure}[!h]
\centerline{\epsfxsize=110mm\epsfbox{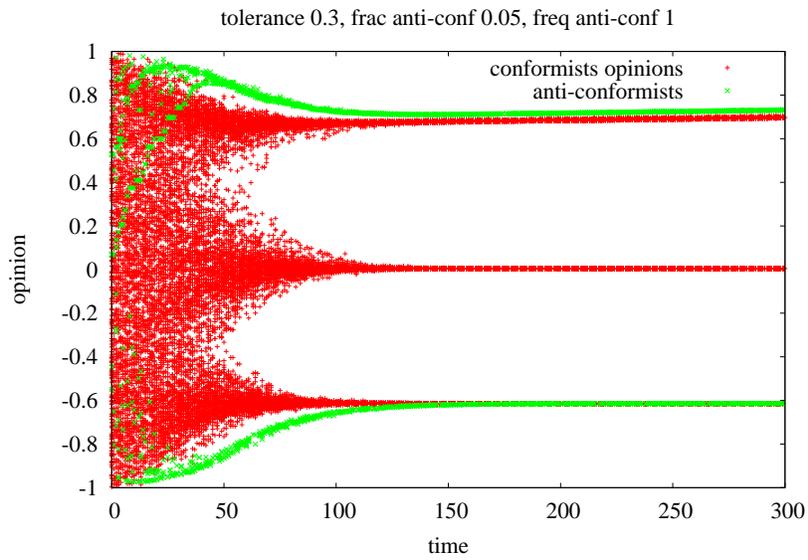}}
\caption{Time evolution of opinions for 0.3 uncertainty level,
$\delta=2$, $\mu=0.1$
and equal chances for opinion expression for conformists (in red)
and anti-conformists (in green). Anti-conformists move to the border
of the distribution in a few tens steps and maintain their
position at the border until convergence of conformists is achieved.} 
\end{figure}

\begin{figure}[!h]
\centerline{\epsfxsize=110mm\epsfbox{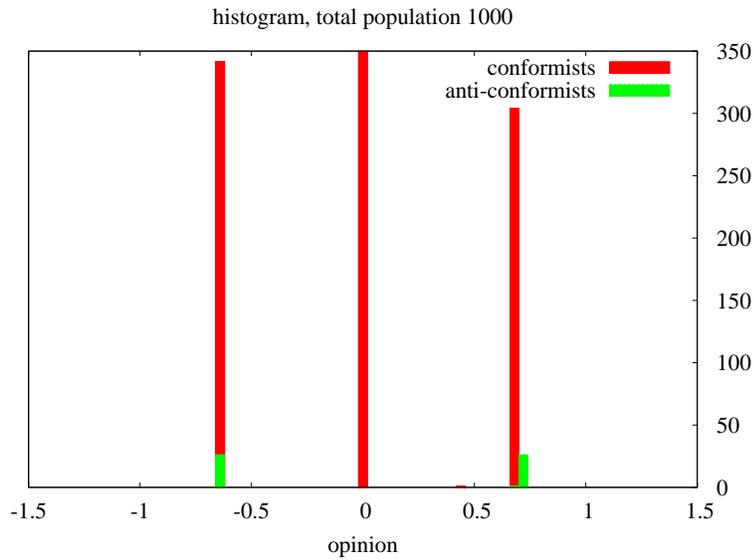}}
\caption{Histogram of asymptotic opinions 
after 300 iterations per individual for 0.3 uncertainty level, $\delta=2$, $\mu=0.1$
and equal chances for opinion expression for conformists (in red)
and anti-conformists (in green) (the same simulation condition as
in figure 5). Apart from the presence of anti-conformists at
the extreme of the distribution, the position of conformists differs little
from the  $int(\frac{1}{u})$ rule prediction, -0.66 and +0.66.} 
\end{figure}

\FloatBarrier

 As a preliminary conclusion, the presence of 5 percent anti-conformists in the population
does not modify the distribution of conformists opinion when anti-conformists 
have the same level of opinion expression as conformists (fig. 5 and 6).

\begin{figure}[!h]
\centerline{\epsfxsize=110mm\epsfbox{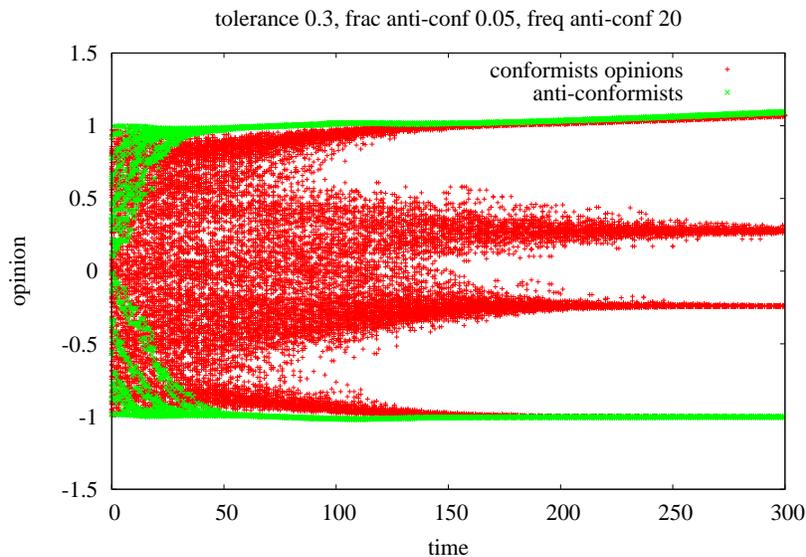}}
\caption{Time evolution of opinions for 0.3 uncertainty level, $\delta=2$, $\mu=0.1$
 when anti-conformists (in green) express 
their views 20 times more often than conformists (in red).
We now observes 4 clusters instead of 3 with quite different positions.
Anti-conformists have attracted two clusters to more extreme positions 
around -1 and +1 and the rest of the conformists have now moved closer to the center.
 } 
\end{figure}

\begin{figure}[!h]
\centerline{\epsfxsize=110mm\epsfbox{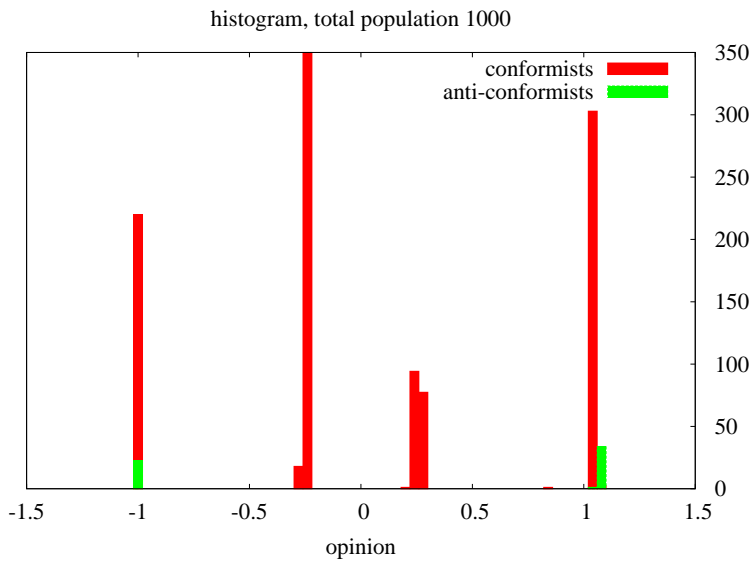}}
\caption{ Histogram of asymptotic opinions for 0.3 uncertainty level, $\delta=2$, $\mu=0.1$ ,
 when anti-conformists (in green)
express their views 20 times more often than conformists (in red).
 We now observes 4 clusters instead of 3 with quite different positions.
Anti-conformists have attracted two clusters to more extreme positions around -1 and +1
and the rest of the conformists have now moved closer to the center.} 
\end{figure}

\FloatBarrier

However as one can observe on the next couple of figures 7 and 8, 
anti-conformists  strongly influence the distribution of
 conformist opinions when anti-conformists 
express their views 20 times more often than conformists.
Figure 7 shows that interaction among conformists make them converge into 4 clusters.
The respective positions of conformists and anti-conformists clusters
results from their mutual interaction: anti-conformists first aggregate
outside conformists, but later, after conformists aggregation, the anti-conformists
 have a reverse motion towards the closest conformists' cluster,
symmetrical to the motion of the extreme conformist cluster towards anticonformists. 


\FloatBarrier

\paragraph{Variability of asymptotic clusters }

  As a matter of fact, several trials with different random sampling 
give qualitatively equivalent results: the presence of a few outspoken anti-conformist
(with $f=20$) changes the attractors of opinion dynamics from
three clusters to two center clusters + two extreme clusters; but 
the cluster positions and amplitude might change noticeably
between simulations with different random samplings for the same set of parameters.
 Such instabilities are well known in random processes
such as Polya urns or Chinese restaurant process (\cite{polya}).
 In the case of Polya urns, a coloured ball is randomly drawn
from the urn and two balls of the same colour are then replaced 
in the urn. When the initial number of balls is small, say one red ball
and one black ball, the proportion of late samplings is strongly dependent upon
 the first occurring draws and is thus susceptible of large fluctuations.
 We are precisely in a similar case, since the initial
opinions and sampling of the 50 anti-conformist have a strong influence on
the outcome of the dynamics. This phenomenon is a particular instance
of the path dependence phenomenon 
observed in non-linear and random processes (\cite{BA,NW}).
Path dependence reflects the influence of history, which importance
in politics is not a surprise to political scientists.

We therefore performed 100 simulations per set of parameters and 
display the results as histograms. In the next 3 figures, red histograms
correspond to conformist opinions at the end of the simulation
and green histograms correspond to anti-conformist opinions.
The blue histograms corresponding to simulations in the absence of 
anti-conformist are shown for the sake of comparison.
Common simulation parameters are anti-conformism strength parameter $\delta=2$,
number of agents 1000, number of iterations per agent 3000,
fraction of anti-conformist 0.05, kinetic factor  $\mu=0.1$
and multiplicative frequency factor of anti-conformist $f=20$.
 They only differ by the uncertainty levels: $u=0.3, 0.4, 0.6$.

\begin{figure}[!h]
\centerline{\epsfxsize=120mm\epsfbox{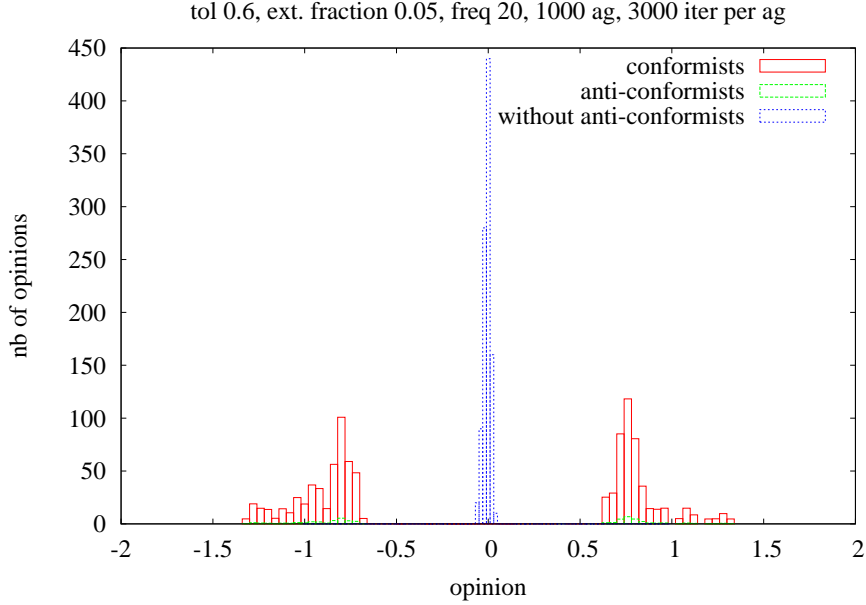}}
\caption{ Histograms of opinions averaged over 100 runs
after 3000 iterations per individual. 0.6 uncertainty level.
The red (resp. green) histograms
are for conformists (resp. anti-conformists) and the blue histogram was obtained for conformists
in the absence of anti-conformists.} 
\end{figure}

\begin{figure}[!h]
\centerline{\epsfxsize=120mm\epsfbox{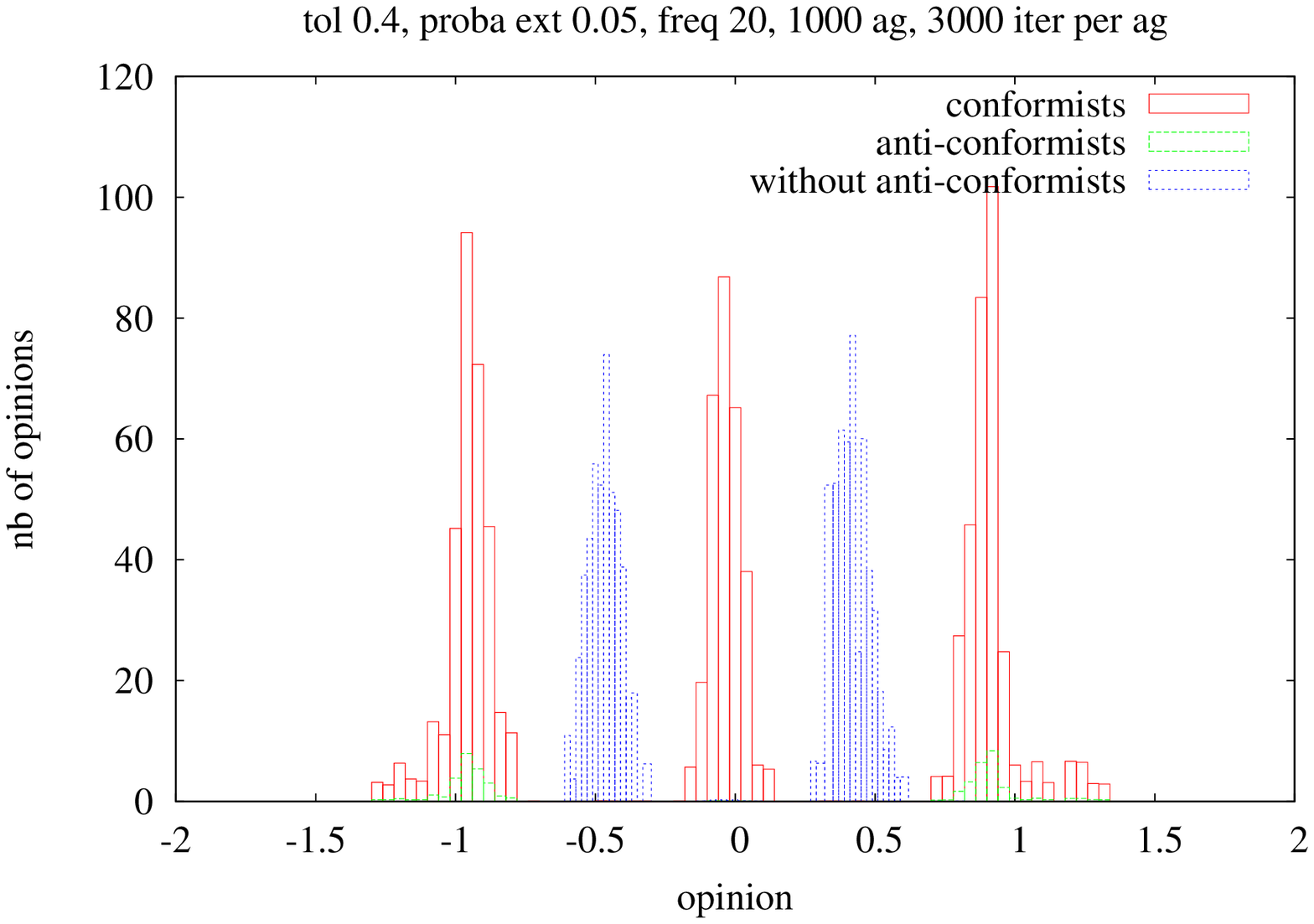}} 
\caption{ Histograms of opinions averaged over 100 runs
after 3000 iterations per individual. 0.4 uncertainty level.
 The red (resp. green) histograms
are for conformists (resp. anti-conformists) and the blue  histogram was obtained for conformists
 in the absence of anti-conformists.} 
\end{figure}

\begin{figure}
\centerline{\epsfxsize=120mm\epsfbox{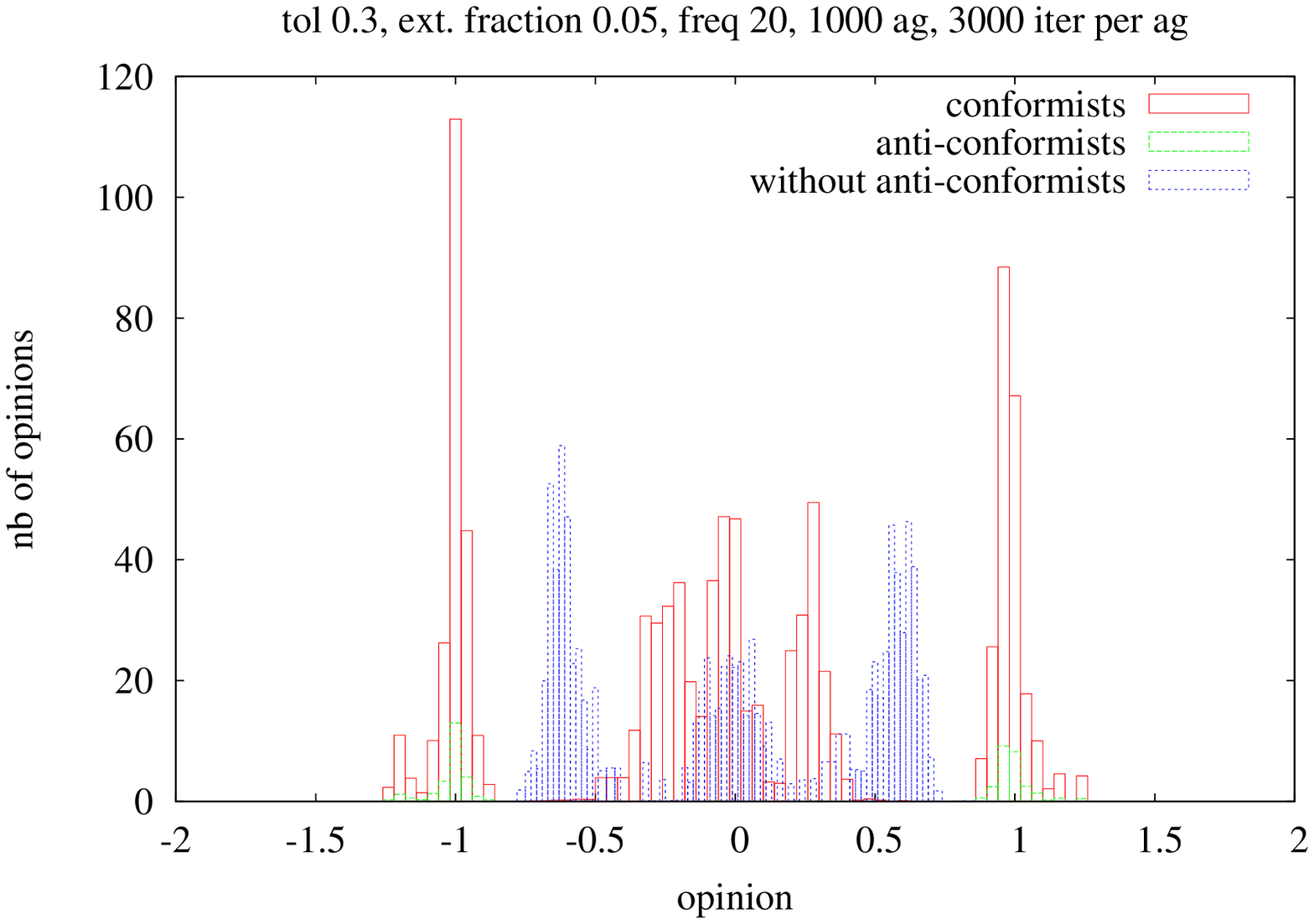}}
\caption{ Histograms of opinions averaged over 100 runs
after 3000 iterations per individual. 0.3 uncertainty level.
The red (resp. green) histograms
are for conformists (resp. anti-conformists) and the blue  histogram was obtained for conformists
 in the absence of anti-conformists.} 
\end{figure}

  The 3 figures confirm that large levels of expression by anti-conformists
allow them to drag significant fractions of conformists towards the two 
emergent extreme clusters, with some variability in positions and
 relative amplitude. The influence of anti-conformists increases the number of clusters
composed of conformists by (at least) one; two extreme clusters have opinions aligned
with those of anti-conformists, while the remaining cluster(s) (which can be 
0, 1 or 2) are centered closer to the original opinion average\footnote{ A word of caution:
 such histograms could be interpreted
 either as histograms of the positions of single isolated
 peaks as observed in figures 5 and 7, or as the aggregation of wider peaks.
 We confirm that only the first interpretation is correct from many direct
observations of asymptotic histograms of single iteration processes.
Furthermore, wide peaks would not be stable under the bounded
confidence process.}.

  The following simple argument explains the increase in the number of clusters
due to the influence of the anti-conformist agents.
Two extreme conformist clusters of initial width $u$ are attracted by the 
anti-conformists. They don't participate in the formation of the  central clusters.
The ``effective'' width $w_c$ of initial segment of conformists which end up
in the central clusters is then reduced to $2-2u$, instead of 2.
 Applying the $n=int(\frac{w_c}{2u})$
rule to the number $n_c$ of central clusters gives $n_c=int(\frac{1-u}{u})$.
 The predicted total number of clusters is then:
 \begin{equation}
   n=2+int(\frac{1-u}{u})
 \end{equation}
  A comparison with simulation results gives:

  \begin{tabular}{|r|c|c|}\hline
    uncertainty & predicted clusters' number & observed number\\\hline
      .6        &         2                &       2        \\\hline
          .4        &         3                &       3        \\\hline
      .3        &         4                &       5 with overlap   \\\hline
  \end{tabular}

\FloatBarrier

\paragraph{Checking the path dependency}

Because the earlier steps of the dynamics are so important,
we might expect that anti-conformists have a stronger influence if they step in
earlier rather than later. In the next two sets of simulations, the frequency factor 
$f$ was either decreased or increased linearly in time between 1 and 20,
for the same set of parameters as in figure 10. One can check that the red histogram
taken for decreasing $f$ from 20 to 1 is nearly the same as the red one on figure 10,
obtained in the presence of 5 perc. anti-conformists with relative expression frequency 20,
while the green one for increasing $f$ from 1 to 20 is nearly the same as the blue one
(obtained in the absence of anti-conformist) on figure 10. In other words, the early expression of 
extremist views determines the outcome of the process, while late expression has 
nearly no effect.

\begin{figure}
\epsfxsize=120mm\epsfbox{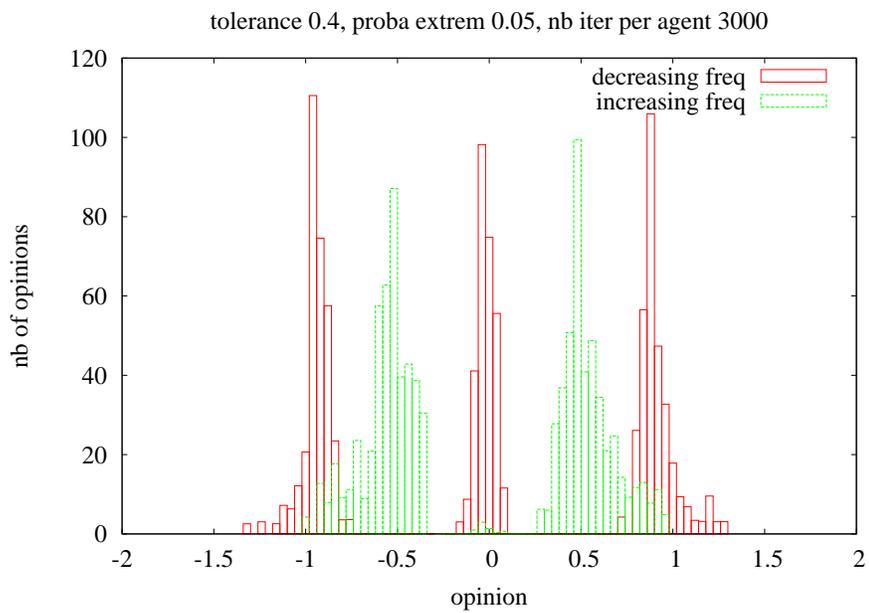}
\caption{ Histograms of conformist opinions 
after 3000 iterations per individual. 0.4 uncertainty level.
The red histogram correspond to decreasing the anti-conformist expression frequency,
the green one to its increase. } 
\end{figure}

\FloatBarrier

\paragraph{The consensus regime}

For larger uncertainty values, such as $u=0.9$, the situation changes
dramatically. Single runs show that there remains only one cluster of conformists at the
same position as the anti-conformist cluster; consensus is restored.
 But a more systematic survey running 300 simulations per set of parameters
shows on figure 13 that the position of the single cluster
 varies widely along the opinion axis; 
furthermore, no peak structure
similar to those observed at lower uncertainties is apparent
 on the histogram of clusters.
 The vertical bars of the histogram
are the number of opinions divided by the number of conformist 
agents. Except for a small region in the neighbourhood of opinion 0.0,
the height of the peaks are integer values, 
indicating that for each simulation,
 all opinions are concentrated in one single cluster.
A first tentative explanation for randomness of the cluster position
rests on the faster dynamics at larger $u$ values: nearly all
pair samplings satisfy the confidence condition for interaction,
and convergence is then faster, as we checked on time plots
(not represented). We have earlier seen that
faster convergence yields more sensitivity to the early steps of
the dynamics and more dispersion of the asymptotic results.

Obtaining a consensus for $u=0.9$ is in accordance with the 
$int(1/u)$ rule of the bounded confidence model; but the 
standard bounded confidence model yields a consensus attractor 
close to the center of gravity of the initial distribution, 
which is quite different from the present result: the consensus peaks 
seem randomly located on the [-1,+1] opinion axis.

 Our results
also differ from those of the extremism model of \cite{daw}
 who predict clusters located close
to the anti-conformist initial positions i.e -1 or +1, for large values of $u$
( in accordance to their hypotheses
of quasi-fixed position of anti-conformist at -1 or +1). By contrast, for large values 
of $u$, in our present model, the 
position of anti-conformist initially  scattered over
the entire $[-1,+1]$ segment results largely from the earlier iteration steps
and can undergo large fluctuations.

  The transition between single cluster dynamics at larger $u$ and
 two clusters dynamics at lower $u$ is smooth; it is a crossover 
rather than a sharp transition
and occurs around $u=0.8$. At $u=0.8$ the histogram (fig 14)
displays co-occurrence of bins with integer values corresponding to single clusters
and of bins with non-integer values clustered in two wide peaks around 
$\pm 0.66$.

\begin{figure}
\centerline{\epsfxsize=120mm\epsfbox{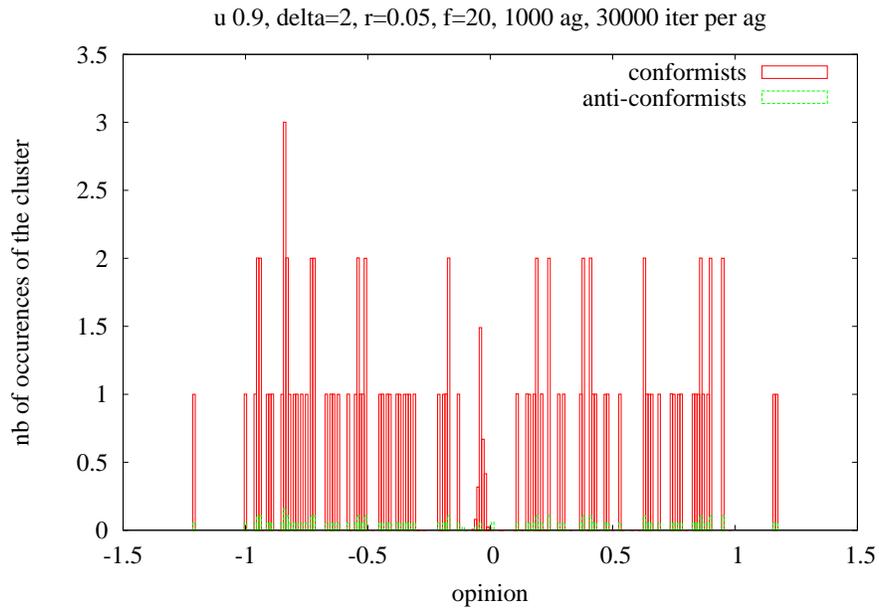}}
\caption{ Histogram of conformist agents asymptotic opinions for a 0.8 uncertainty level
based on 300 simulations.
 The vertical bars
are the number of opinions divided by the number of conformist 
agents. 30000 iterations per individual for a 0.9 uncertainty level.
Except for a small region in the neighbourhood of opinion 0.0,
the height of the bins are integer values, 1 ,2 or 3 
indicating that for each simulation,
 all opinions are concentrated in one single cluster} 
\end{figure}

\begin{figure}
\centerline{\epsfxsize=120mm\epsfbox{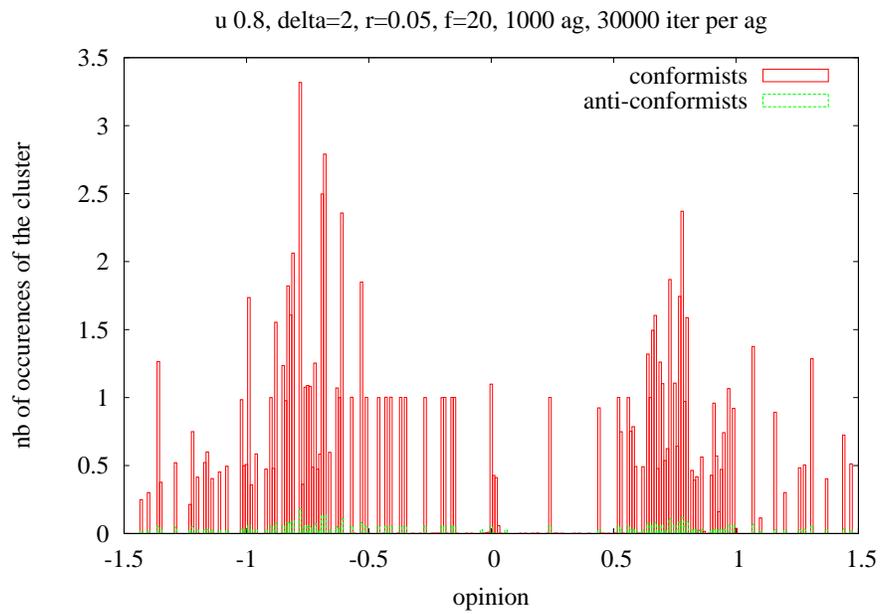}}
\caption{ Histogram of conformist agents asymptotic opinions for a 0.8 uncertainty level
based on 300 simulations. 3000 iterations per individual for a 0.8 uncertainty level, 
corresponding to the transition. The bins are a mixture of integer values
1 and 2, plus two wide peaks around $\pm 0.66$ } 
\end{figure}
\FloatBarrier

\paragraph{Influence of simulation parameters}

In order to study the influence of the different
parameters $r, f, u, \mu $ and $\delta$ on the outcome of opinion dynamics,
one has to compress the information in the histograms by monitoring
some of their characteristics. We have chosen to monitor the characteristics
of the positive 'extreme' peak, the rightmost peak\footnote{Isolating the 
 rightmost peak for these measurements was done by checking the histograms
for a gap left of the peak and taking measurements on the remaining
bins right of the gap; for figure 10 e.g., the empty bin at 0.65 opinion 
can be used to start collecting the statistics.} 
  observed on figures 9, 10, 11.
 We monitored
 the fraction of opinions in the peak
(i.e how many conformists were attracted by the rightmost anti-conformists), 
their average deviation (how far from the initial average 0 they were attracted),
 the standard deviation 
of the distribution of opinions in the peak\footnote{ These three quantities correspond to 
standard measurements of peak characteristics in spectra, the area under the peak (fraction),
the peak position with respect to the origin (average deviation), and the peak width
 (twice the standard deviation). For figure 10 e.g. the fraction of opinions
in the righmost peak is 33 perc., the average position is 0.91 and the standard deviation 0.10. },
 and finally the product 
of the average times the fraction\footnote{In the next five figures,
the fraction of opinions, the standard deviation and the attractiveness
are given by the scale on the left, in red, and the average 
deviation by the scale on the right, in green.}
. This latest quantity measures
some kind of ``attractiveness''. We shall see that this attractiveness
often (but not always as further discussed)
 displays relatively little variation, corresponding to a balance
between how many conformists are attracted by anti-conformists and how far they are attracted. 

 The present systematic investigation of the role of simulation parameters is
limited to the lower values\footnote{larger values of $u$ were previously investigated
in the section on the consensus regime.} of $u$, in the multiple clusters regime.

\begin{figure}[!h]
\centerline{\epsfxsize=120mm\epsfbox{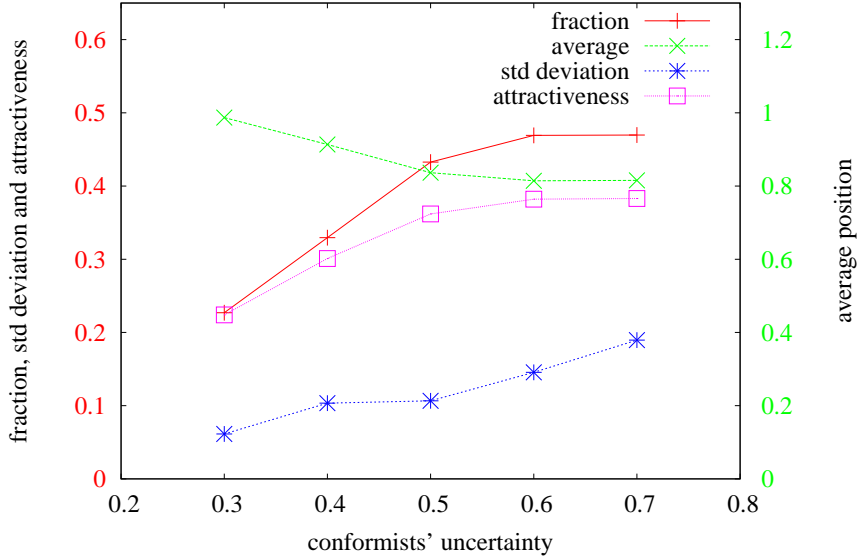}}
\caption{ Variations of the rightmost peak characteristics
 with uncertainty $u$.
The fraction of opinions, the standard deviation and the attractiveness
are given by the scale on the left, in red, and the average deviation by
 the scale on the right, in green.
When uncertainty increases
from  $u=0.3$, to $u=0.7$,
 the fraction of attracted conformists (the red crosses)
increases up to values (0.47) corresponding to a near depletion of
 the central peak(s) where only 6 perc. of the population remains;
most conformists became extremists.
 Average opinion (green crosses) of the extreme peak decreases to 0.8 and
its standard deviation increases.
}
\end{figure}

Applying this method to the {\bf influence of uncertainty $u$}
 one obtains the following graph (fig.15), some points ($u=0.3, 0.4, 0.6$)
 of which can be checked against the histograms on fig. 9, 10 and 11.
 The linear increase with $u$ of the fraction of agents attracted to the extreme peak
is simply understood from the width of the conformist's zone under the influence
of the anti-conformist: since anti-conformist move early to the upper boundary of the conformist
cluster they can at most influence a fraction $u$ of them. The simulated results
give fractions of 0.227 for $u=0.3$, 0.33 for $u=0.4$ and 0.43 for $u=0.6$, 
not far away from the $u$ upper-bound.
But the attracted fraction saturates close to 0.5
 at larger $u$ values, when the two extremist 
clusters are competing.


Increasing the {\bf fraction $r$} of anti-conformist (fig. 16) and
 their relative {\bf frequency of intervention $f$} (fig. 17) 
 increases the average deviation of the peak from 0.
The ratio of biased moves of conformists towards anti-conformists to their converging moves
obtained from the tree of probabilities (figure 4) is $\frac{rf}{1-r}$
This ratio increases with both $r$ and $f$, and so does the average deviation of the peak.

\begin{figure}[!h]
\centerline{\epsfxsize=120mm\epsfbox{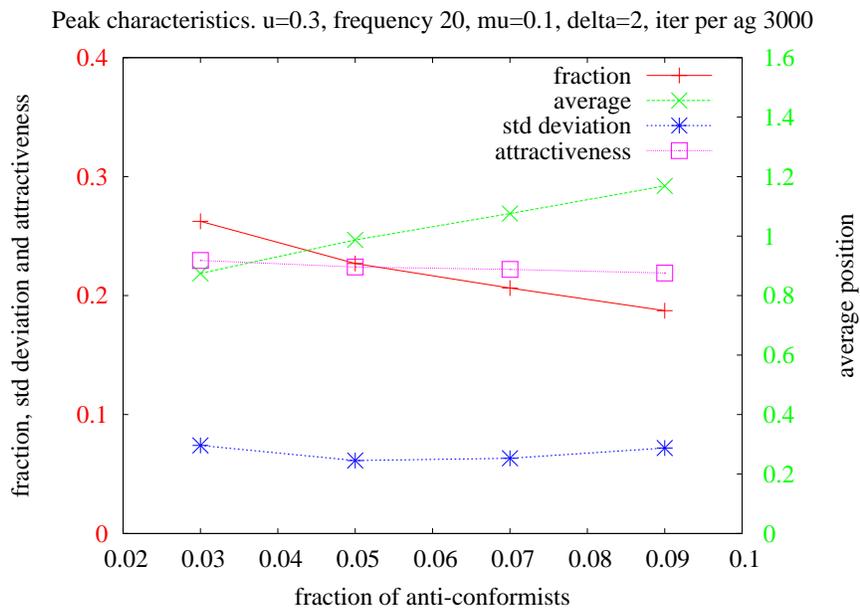}}
\caption{ Variations of the rightmost peak characteristics with $r$
the fraction of anti-conformists.
The fraction of opinions, the standard deviation and the attractiveness
are given by the scale on the left, in red, and the average deviation by
 the scale on the right, in green.} 
\end{figure}

\begin{figure}[!h]
\centerline{\epsfxsize=120mm\epsfbox{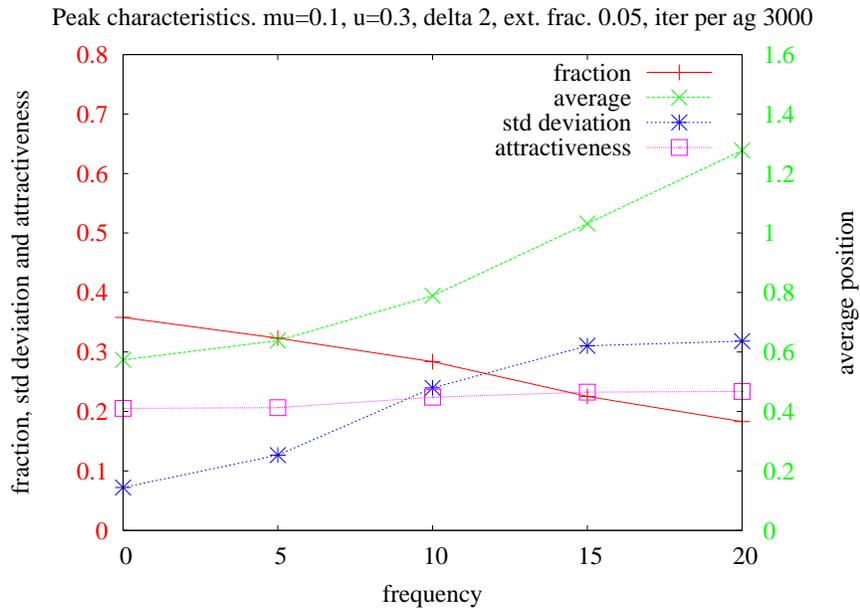}}
\caption{ Variations of the rightmost peak characteristics with $f$
the multiplicative factor of interactions with anti-conformist.
The fraction of opinions, the standard deviation and the attractiveness
are given by the scale on the left, in red, and the average deviation by
 the scale on the right, in green.} 
\end{figure}

\FloatBarrier

  {\bf $\mu$} is a priori a simple {\bf kinetic parameter} which increase
 reduces the convergence time.
For instance, a value of 0.5 was generally considered as optimal
in bounded confidence models since it implies the full agreement
between a pair of agents
 on the middle position in one single iteration.
But as observed with this plot (fig. 18),
  $\mu$ also influences the peak characteristics.
When convergence is fast, the initial steps of the iteration process
have an even stronger influence on the outcome of the dynamics - see
for instance the dramatic increase of the standard deviation
of the rightmost peak when $\mu = 0.4$.
 A technical conclusion is that in order to avoid
strong sampling variations, opinion dynamics models should be run with values of
 $\mu \leq 0.25$. This is anyway compatible with the fact that 
in real life several interactions are necessary to significantly change
opinions.

\begin{figure}[!h]
\centerline{\epsfxsize=120mm\epsfbox{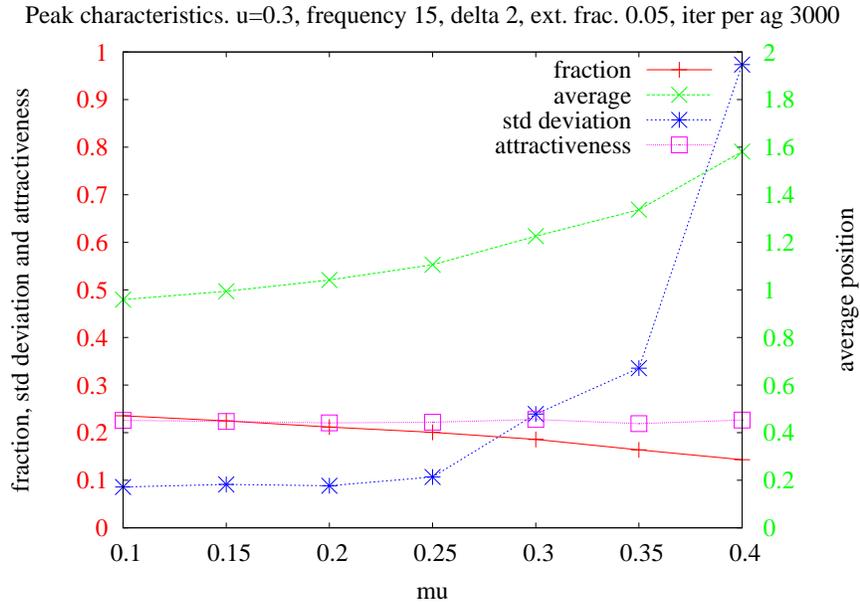}}
\caption{ Variations of the rightmost peak characteristics with $\mu$
the kinetic factor.
The fraction of opinions, the standard deviation and the attractiveness
are given by the scale on the left, in red, and the average deviation by
 the scale on the right, in green.}  
\end{figure}

\FloatBarrier

    The variations of the rightmost peak characteristics with {\bf $\delta$
the anti-conformism intensity} are non-monotonic. When $\delta$ increases
from 1.5 to 2.6, the average deviation increases, 
which is a direct consequence of equation (2).
But it reaches a maximum
around $\delta=2.6$.
The change of slopes of the average deviation, of the fraction of conformists in the peak and of 
the attractiveness curves, as the strong increase of the standard deviation
are evidences of a regime transition around $\delta=2.6$

\begin{figure}
\centerline{\epsfxsize=120mm\epsfbox{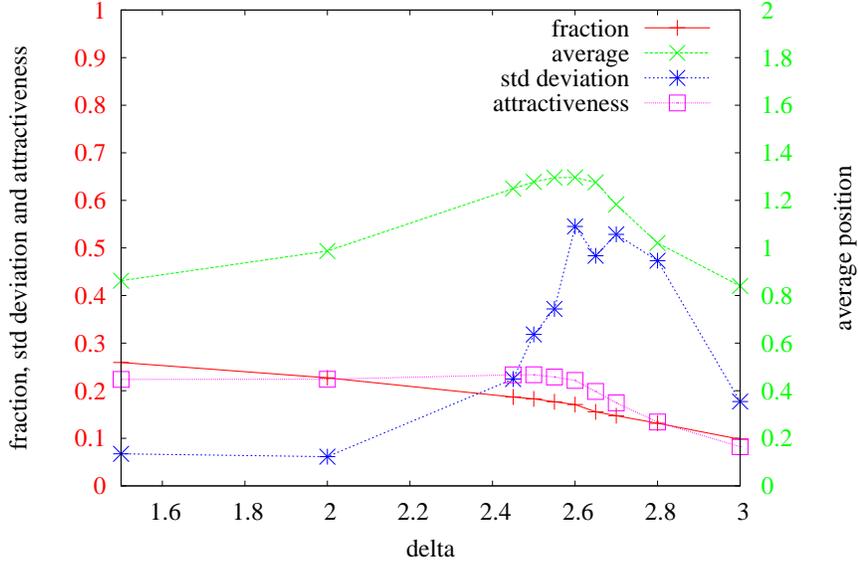}}
\caption{ Variations of the rightmost peak characteristics with $\delta$
the anti-conformism intensity. When $\delta$ starts increasing
from  $\delta=1$, the conformists are attracted to more extremism (the green
points). But when $\delta=2.6$ a regime changes occurs as observed on
all four monitored quantities: extremists' fraction (red crosses)
and attractiveness (pink squares)
 decrease, and they start loosing followers (red crosses).
The strong increase of the standard deviation around $\delta=2.6$
is also a clue of the regime transition. 
} 
\end{figure}

\FloatBarrier

  To observe the transition region in detail (fig. 20),
we came back to individual asymptotic histograms
similar to figures 6 and 8.
 One notices that before the transition, at $\delta=2.4$ the leftmost
and rightmost peaks of conformists histograms (in red) occur at the same opinion values as
anti-conformist histogram peaks (in green). For higher values of $\delta$
the anti-conformist peaks are outside the leftmost
and rightmost conformists peaks. In other words, the anti-conformists
became unable to drag anymore the conformists to their position. 

Since larger values of conformists uncertainty $u$ make them more susceptible
to the influence of anti-conformists, the stalling
transition value $\delta_s$ increases with $u$: we observed 
a transition at $\delta_s=2.5$ when $u=0.3$, at $\delta_s=2.8$ when $u=0.4$
and at $\delta_s=3.2$ when $u=0.6$.

\begin{figure}[!h]
\centerline{\epsfxsize=120mm\epsfbox{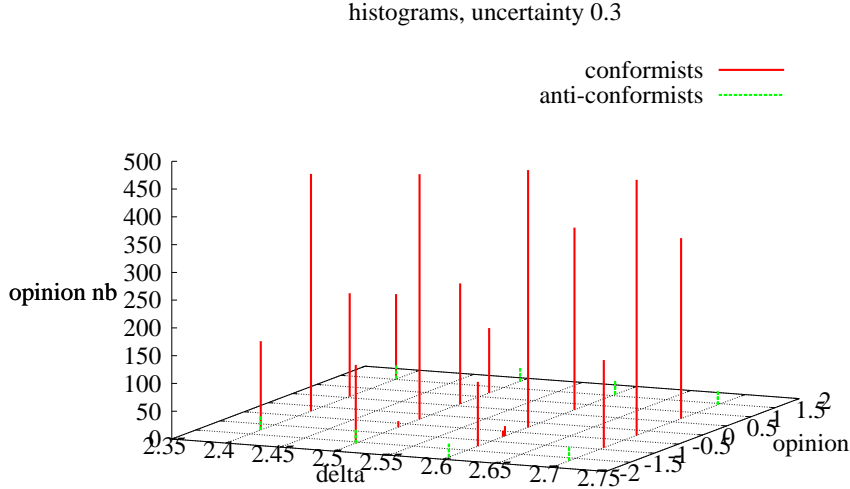}}
\caption{ The stalling transition: asymptotic opinion histograms for $\delta= 2.4, 2.5, 2.6, 2.7$.
 The red (resp. green) histograms
are for conformists (resp. anti-conformists). Extreme peaks coincide in position
at $\delta=2.4$, but they start diverging at $\delta=2.5$ }  
\end{figure}

\FloatBarrier

 To summarise on ``Attractiveness'', it displays relatively 
small variations with $r,f$ and $\mu$, reflecting a rough balance\footnote{we can 
only conjecture about this balance, but we have no explanation for it,
nor why changes in attractiveness or its derivatives goes along with
 transitions in dynamical regimes.}
 between how many
conformists are attracted toward extremism and how far.  The same relative stability
is observed with respect to $\delta$ until $\delta=2.5$. Attractiveness decays when 
$\delta>2.5$ and this is an indication of a change in dynamical regime:
anti-conformists lost their strong influence on conformists. The increase in 
``attractiveness'' with conformists's uncertainty $u$ also reflects the series
 of transitions in the number of clusters with $u$.

\section{Discussion and conclusions}

 Let us first summarise our results.

\begin{itemize}

\item Anti-conformism of small fraction of the agents population
can result in the emergence
of large extremist clusters, provided that the 
anti-conformists express more often their views than conformists.  

\item This influence exists whatever conformist uncertainty,
and it is larger when uncertainty increases. Two distinct dynamical regimes
are observed according to the value of uncertainty. 
For lower values, anti-conformists drag important fractions of conformist 
agents to their own extreme position. For higher values of uncertainty,
consensus is restored, but along a much wider range of positions 
which can be centered far away from the initial center of gravity of
initial opinions.   

\item Obviously the anti-conformist influence increases with their
number and the frequency of their interventions.
By contrast, one observes a transition in the anti-conformist
influence when anti-conformists position themselves too far away from
the center; they then loose influence and are unable to drag large 
fractions of conformists.

\item Early intervention of anti-conformists increases their influence.
And the early steps of the dynamics are responsible for the large
deviations in peak positions.

\item
  The results concerning the number of peaks in the opinion distribution
 as a function of the uncertainty parameter and their
approximate position are robust. The exact position of the 
peak cannot be predicted accurately, due to the susceptibility
of the probabilistic dynamics to initial samplings.

\end{itemize}

  Let us now discuss how these conclusions would eventually
be modified by the other players in the political game,
media, parties, and other institutions such as elections, government etc.

In fact, media and political parties re-enforce the influence
of non-conformists. Journals, newspapers or television
compete for readership and audience. Journalists fight for impact, notoriety
and reputation. In this market for information, or cognitive market as 
proposed by \cite{bron}, the motivations are the same as for anti-conformists
of \cite{smalep}. Impact is achieved by taking simple, extreme and fast
positions. The tendency is increased by the use of Internet,
from which journalists often take their views. The fast communication
procedures on Social Networks also favour the extremes as observed
on tweets and readers reaction to articles in the press. To maximise 
audience, societal and political debates on the television are dramatised:
they oppose extreme views and seldom result in consensus.
 As a matter of fact, the media contribute largely to the high value
of the relative frequency factor $f$ used in our simulations.
 In that respect, the growing role of the media and especially
of the Internet will not automatically lead to a better understanding
of challenges and options, but might on the contrary favour
the expression of extremist views.   

 The same mechanisms can be observed during the political debate
inside parties, before elections. Party members are competing to get
positions inside the party or to represent the party in future elections. 
They also want to
make clear that they are faithful to their party by strongly opposing
other parties views. For them too, a simple ideological position
is easier to express and to defend, than balancing between the contradictory
constraints faced in the choice of a policy adapted to societal challenges.

So both media and political parties internal discussions re-reinforce the 
influence of extremists.

 The dynamics might be different during elections 
and at the government level. 
On the occasion of national elections for instance, parties have
to adapt their program to the electorate and make alliances to
win support. In principle they should move towards the center 
for this. But when the electorate comprises strong extremist clusters,
as often observed in our simulations,
they have a choice to position themselves clearly on one side of the
political checkers, especially under the influence of their members
which are biased with respect to the 'rational' position of optimising support
from the general population.

 The government itself has to navigate between general support
and the support from inside the parties of the alliance which
brought it in power.

  In conclusion of the present discussion, the dynamical
processes inside the media and the parties are in agreement with
our hypothesis of a stronger expression of anti-conformist
positions. This re-enforce the conclusions of our model.

On the other hand, other aspects of politics concerning general elections
or government positions necessitate further analysis.

What we tried to demonstrate is that evolution towards extremism
does not automatically imply coercion, strategic plots or the
control of the media by a single agent. Simple human cognitive processes
such as anti-conformism, cognitive biases and uncertainty of agents
can favour its emergence and its influence on the constituency.  

  The results of our simulations were interpreted in terms of
politics, but they could also provide some insight into
other social phenomena involving the dynamics of extreme choices:
\begin{itemize}
\item
In markets of luxury goods: for instance, why do people buy fast cars or SUV vehicles
when they have little use for these products? How is the market
driven by these extreme choices?
\item in Fashion and in the Arts, where anti-conformism is the rule driving
the perpetual motion of expressed realisations; 
\item in the propagation of imaginary dangers related to new technologies in the
media and the Internet (\cite{bron}).
\end{itemize}

{\bf Acknowledgements}
 We thank Joshua Epstein and Paul Smaldino for sending their preprint
prior to publication and the participants to the ``s\'eminaire inattendu''
for their comments. We thank anonymous referees for their corrections and 
for raising interesting issues.

\bibliography{biblio}

\end{document}